%% file: main.tex
\documentclass[conference]{IEEEtran}

\usepackage{nameref}
\usepackage{braket}
\usepackage{qcircuit}
\usepackage{cite}
\usepackage{amsmath,amssymb,amsfonts}
\usepackage{algorithmic}
\usepackage{graphicx}
\usepackage{textcomp}
\usepackage{xcolor}
\usepackage{mathptmx} 
\def\BibTeX{{\rm B\kern-.05em{\sc i\kern-.025em b}\kern-.08em
T\kern-.1667em\lower.7ex\hbox{E}\kern-.125emX}}

\begin{document}


\title{Conference Paper Title*\\
{\footnotesize \textsuperscript{*}Note: Sub-titles are not captured in Xplore and
should not be used}
\thanks{Identify applicable funding agency here. If none, delete this.}
}

\author{\IEEEauthorblockN{Danylo Lykov}
\IEEEauthorblockA{\textit{Computational Science Division} \\
\textit{Argonne National Laboratory}\\
Lemont, IL 60439, U.S.A. \\
dlykov@anl.gov}
\and
\IEEEauthorblockN{Yuri Alexeev}
\IEEEauthorblockA{\textit{Computational Science Division} \\
\textit{Argonne National Laboratory}\\
Lemont, IL 60439, U.S.A. \\
yuri@anl.gov}
}

\title{Importance of Diagonal Gates in Tensor Network Simulations}

\date{\today} 

\maketitle

\begin{abstract}

In this work we present two techniques that tremendously increase the performance of 
tensor-network based quantum circuit simulations. The techniques are implemented in the QTensor package and benchmarked using Quantum Approximate Optimization Algorithm (QAOA) circuits. 
The techniques allowed us to increase the depth and size of QAOA circuits that can be simulated. In particular, we increased the QAOA depth from 2 to 5 and the size of a QAOA circuit from 180 to 244 qubits. Moreover, we increased the speed of simulations by up to 10 million times. Our work provides important insights into how various techniques can dramatically speed up the simulations of circuits.

\end{abstract}

\section{Introduction}
Quantum information science has a tremendous potential to speed up calculations of certain problems over classical calculations \cite{alexeev2019quantumworkshop}.
To continue the advances in this field, however, often requires  classically simulating quantum circuits. Such simulation is done by using classical simulation algorithms that replicate the behavior of executing quantum circuits on quantum hardware on classical hardware such as personal computers or high-performance computing (HPC) systems.
These algorithms play an important role and can be used to (1) verify the correctness
of quantum hardware,
(2)  help the development of hybrid classical-quantum algorithms,
(3) find optimal circuit parameters for hybrid variational quantum algorithms,
(4)  validate the design of new quantum circuits, and (5)  verify quantum supremacy and advantage claims.

Several approaches have been employed to simulate quantum circuits.
The major types include the state-vector evolution approach \cite{de2007massively, smelyanskiy2016qhipster, haner20170, wu2019full, wu2018amplitude, wu2018memory},
linear algebra open system simulation \cite{quac},
and tensor network contractions \cite{markov2008simulating, pednault2017breaking, boixo2017simulation, lykov2021large}.
All these simulators have various advantages and disadvantages. 
For example, the state-vector evolution approach, while being relatively easy to implement,
has an exponential memory requirement with respect to the number of qubits in the circuit, which is a major bottleneck preventing the quantum simulations beyond approximately 46 qubits on modern supercomputers.

In our opinion the most promising type of simulator is the  tensor network contraction approach. It is especially efficient for simulating shallow-depth circuits.
This approach can be sensitive to the connectivity of a quantum circuit and the types of gates.
In this paper we describe the tensor network simulator implementation and show two optimization techniques 
that enable dramatic speedup of simulations. We use quantum circuits from the Quantum Approximate Optimization Algorithm (QAOA) algorithm since it is a 
promising candidate for  demonstrating quantum advantage and benchmarking quantum devices.

All simulations in our paper used  QTensor \cite{qtensor}, developed at  
Argonne National Laboratory. It is a quantum circuit simulator that
uses a tensor network contraction approach with a special 
focus on the simulation of QAOA circuits. It supports simulating both probability amplitudes
and energy expectation values.

In the following section  we introduce the tensor network contraction approach and describe the 
QAOA quantum circuits. In particular, we show how the usage of the Feynman path formalism provides the 
possibility for optimization.
We then describe the optimization techniques and the resulting speedup of simulations.
The final section contains our conclusions and further directions of research.


\input{inputs/graph_figures}

\section{QAOA algorithm}
The QAOA algorithm, introduced by Farhi and Goldstone in 2016 \cite{farhi2016quantum}, 
is a seminal hybrid quantum-classical algorithm for approximate optimization.
The algorithm can be used to find approximate solutions to NP-complete
combinatorial optimization problems.
Here we will demonstrate how it works to solve the MaxCut problem and all benchmarking simulations are performed for MaxCut, but this work also applies to other types of tensor network simulations.

The goal of the MaxCut problem is to color all vertices of a given graph $G=(V,E)$,
such that the two resulting parts of the graph are connected to each other
by the maximum number of edges.
The cost function for MaxCut is $C= \frac{1}{2} \sum_{<jk>\in E} -Z_j Z_k + 1$,
where $Z_i$ is a label of each vertex.
To solve this problem using QAOA, one has to find optimal parameters
for the parametrized ansatz state $\ket{\gamma\beta}$ such that the expectation value
of the Hamiltonian cost function $\braket{\gamma\beta|\hat H | \gamma \beta}$ is maximized.
The ansatz state depends on two parameter vectors $\gamma$ and $\beta$.
The length of the parameter vector, denoted $p$, is an important parameter
that defines the quality of the solution.
For a QAOA depth $p$ with MaxCut on graph $G=(V,E)$, the ansatz state is equal to

\begin{equation}
\ket{\gamma \beta}_p = \prod_{k=1}^p U_C(\gamma_k)U_B(\beta_k) \ket + ,
\label{eq:qaoa1}
\end{equation}
where $U_B(\beta) = e^{-\imath\beta\sum_{j\in V}X_j}$
and $U_C(\gamma) = e^{-\imath\frac{\gamma}{2}\sum_{(i,j)\in E}(I - Z_iZ_j)}$.
Substituting $U_B$ and $U_C$ into (\ref{eq:qaoa1}) and discarding
the global phase, we  obtain
\begin{equation}
     \ket{\gamma\beta}_p =
     \prod_{q=1}^p
     \ [ 
     \prod_{ij\in E, k \in V} e^{\imath\gamma_qZ_iZ_j}  e^{-\imath\beta_q X_k}
     ] |+\rangle .
\end{equation}
A quantum circuit that generates the ansatz state for MaxCut on a fully connected 4-node graph is 
shown in Figure \ref{fig:circ}. For a more detailed description of QAOA see \cite{farhi2016quantum}.

The optimal $\gamma, \beta$ parameters
correspond to the minimum of the Hamiltonian cost function expectation value,
which can be calculated as $E = \bra{\gamma \beta} C \ket{\gamma\beta}$.
Note that $C$ is a sum of $|E|$ elements, where each element corresponds to
an edge of the original graph on which we solve MaxCut.
Hence, the expectation value is $E = \sum_k E_k$, where each element is a
matrix element of a local operator that acts on two qubits.
This locality gives room for efficient optimization by canceling 
all the conjugate gates that commute through that local operator.
This optimization,  called lightcone optimization,  was introduced by Farhi and Goldstone in \cite{farhi2016quantum}.

The energy calculation is an important part of the QAOA method, since
one can use a classical computer to optimize the $\gamma, \beta$ parameters
without having to use noisy quantum devices.
In this work, however, we focus on simulating a single amplitude of the
ansatz state as a benchmark for demonstrating the optimizations.
The results for energy simulations will be discussed in future works.

\section{Methodology}
\label{sec:meth}

\subsection{Tensor network approach}
Quantum computers operate by applying gates to 
a quantum state that describes a quantum system consisting of $N$ subsystems
(qubits).
One way to describe the
evolution of the quantum system
is to apply quantum gates in  matrix form on the wavefunction in the form of a state vector.
With the state-vector evolution approach, 
each gate action should be described as an operator that 
acts on the whole system, even if the gate is local to a certain subsystem of qubits. As a result, the whole state vector needs to be stored, which is extremely inefficient in terms of computational resources and memory.

The tensor network approach associates a state vector of the system
with a tensor of $N$ indices. Each index in this tensor
labels the state of a particular subsystem. That is,
the dimension of the index is equal to number of states of the subsystem,
which is always 2 in our case.
Each gate that acts on a subsystem can then be described as
a tensor.
The amplitudes of the resulting state can be calculated by summing the product of the state tensor and 
the operator tensor over the index of the subsystem.

For example, given a system of two qubits and operator $\hat X_0$ acting 
on the first qubit, the resulting state in the state-vector formalism would be 
$\ket \phi = \hat X_0 \otimes \hat I_1 \ket \psi$.
In tensor network representation, this equation is   $\phi_{i'j} = X_{i'i}\psi_{ij}$.

If the system is in a product state, then the corresponding state-tensor
is a product of smaller tensors representing each subsystem state. In particular,
a 2-qubit system in a state $\ket 0$ is represented by 
$
\psi_{ij} = 
\begin{pmatrix}
1\\0
\end{pmatrix}_i
\begin{pmatrix}
1\\0
\end{pmatrix}_j
$.
Note that the size of this object is $2N$ compared with that of $2^N$ in the 
state-vector notation. More details on tensor network formalism are available in  \cite{cichocki2016tensor}.

\subsection{Tensor network contraction}

Simulation of probability amplitudes
of a quantum state can be done by contracting a tensor
network that represents the quantum circuit that generates the state.

The contraction of a general tensor network can be written by using a
line graph approach as following:

\begin{equation}
    R_{i_1, \dots, i_p} = \sum_{j_1, \dots, j_{q}}
    \prod_{e_i\in F} W^{i}_{e_i},
    \label{eq:tn}
\end{equation}
where tensor indices $i_{...}, j_{...} \in U$ are represented by
vertices of a hypergraph $L=(U, F)$,
$e\in F$ is a hyperedge of $L$, 
edges are tuples of indices
$e_i=(v_1, v_2, \dots, v_d), \forall k \ v_k \in U$, and tensors $W^i_{e_i}$ have
the number of dimensions~$d$ , the same as the number of vertices in a corresponding edge.
For two-level quantum systems, where each tensor 
dimension has size 2, the sum (\ref{eq:tn}) has $2^{q}$ elements, and each element
corresponds to assignment of $0,1$ to each variable.

Instead of calculating every element of this sum,
one can merge tensors with each other, producing an intermediary tensor
after each merge operation. One way to do this is by selecting some
vertices~$j_i$ from all contracted vertices~$\{j_k\}_{k=1}^q$
and evaluating values of a new intermediary
tensor by summing over $j_i$ a product of only those tensors
that have $j_i$ as their index.
The order in which the $j_i$ are selected determines
the size of the largest tensor that needs to be stored 
as an intermediate step in the contraction.
Thus the total contraction speed and memory 
requirements are determined by this intermediate largest tensor, which can reach
a very large number of dimensions.
More information on 
ordering tensor networks is available in  \cite{schutski2019adaptive}, \cite{lykov2021large}, \cite{detcher2013bucket}.

\section{Optimization techniques}

To speed up and reduce memory requirements of tensor network contraction, we applied two techniques, which we  describe in detail in this section. In the first technique we combined gates, and in the second technique we took advantage of the diagonal properties of the gates.

\subsection{Optimization of QAOA circuit structure}
\label{sec:zzgate}

A typical QAOA circuit is shown in Figure \ref{fig:circ}. We note that Pauli-Z gates are enclosed by two CNOT gates or by Hadamard gates. One can merge these  gates in a 
specialized 2-qubit gate with a parameter~$\gamma$ or parameter~$\beta$ respectively, as shown in the equations below.

\input{inputs/ZZ_replace}
\vspace{0.5em}

\begin{equation}
\hat {ZZ} = e^{i\gamma\hat Z_i \hat Z_j}
\end{equation}

This gate optimization technique reduces the complexity of the tensor network line graph, as shown in Figure \ref{fig:line_graphs}. It also makes finding an optimal tensor contraction sequence easier since the line graph has fewer vertices.

\subsection{Diagonal gate simplification}
\label{sec:diagonal_gates}

A certain property of tensors $W^{i}_{e_i}$ can 
provide an opportunity for optimization in terms of how these tensors are stored and computed.
Each index of sum (\ref{eq:tn}) can have a value $0$ or $1$ for an $N$-qubit system, 
and each assignment of values to indices corresponds to
a single Feynman path which evaluates to an element of the sum (\ref{eq:tn}). 
Since the value of each Feynman path is a product of 
values of different tensors, we know in advance that if for some assignment the value of any tensor is 0, the whole contribution is 0 as well.

In particular, if a tensor $W^{i_0}$ from  Equation \ref{eq:tn} is diagonal, in other words  
$W^{i_0}_{lm} = \alpha_l\delta_{lm}$, then
for any 
assignment of indices~$(l,m) = e_{i_0}$, $l,m\in U$ from sum (\ref{eq:tn}) in which values of
the diagonal tensor indices match,
the corresponding element in the sum will be equal to zero.
The tensor~$W^{i_0}$ can then be safely removed from 
the tensor network and replaced by $\alpha_l$ without changing the result.

Here is a 2-gate example demonstrating how our diagonalization technique is applied.

\begin{equation}
\label{eq:example}
\begin{split}
    |\phi\rangle &= \hat U\hat D |\psi\rangle
    \phi_i = \sum_{jk} U_{ij}D_{jk} \psi_k
    \\
    &=
    \sum_{jk} U_{ij}\alpha_j \delta_{jk} \psi_k = 
    \sum_{j} U_{ij}\alpha_j \psi_j
\end{split}
\end{equation}

We note that  QAOA circuits have only one type of
a 2-qubit gate: $\hat{ZZ} = e^{i\gamma\hat Z_i \hat Z_j}$.
The $\hat Z$ gate is diagonal, as is $\hat Z_i\hat Z_j$;
therefore, the matrix of the $\hat{ZZ}$ gate in the 2-qubit basis will be
diagonal.

\begin{equation}
    \hat {ZZ(\gamma)} =
    e^{i\gamma\hat Z_i \hat Z_j} = 
    \text{diag}(e^{i\gamma},
    e^{-i\gamma},
    e^{-i\gamma},
    e^{i\gamma}
    )
\end{equation}

We can use this fact to replace the 4-index tensor~$U_{ijkl}$ representing
a generic 2-qubit gate 
with a 2-index tensor $U_{ij}=(\begin{smallmatrix}\rho & \overline \rho
\\ \overline \rho & \rho\end{smallmatrix})$,~$\rho = e^{i\gamma}$
that represents a diagonal gate.
This significantly reduces the computational cost for tensor contraction and the memory requirements to store these tensors.

\section{Results}
\label{sec:results}

\newcommand{\nseeds}{{5 }}

\newcommand{\shadeddescr}{{
The shaded region shows 1-$\sigma$ interval over \nseeds random graphs.
}}

\begin{figure}
\centering
\includegraphics[width=\linewidth]{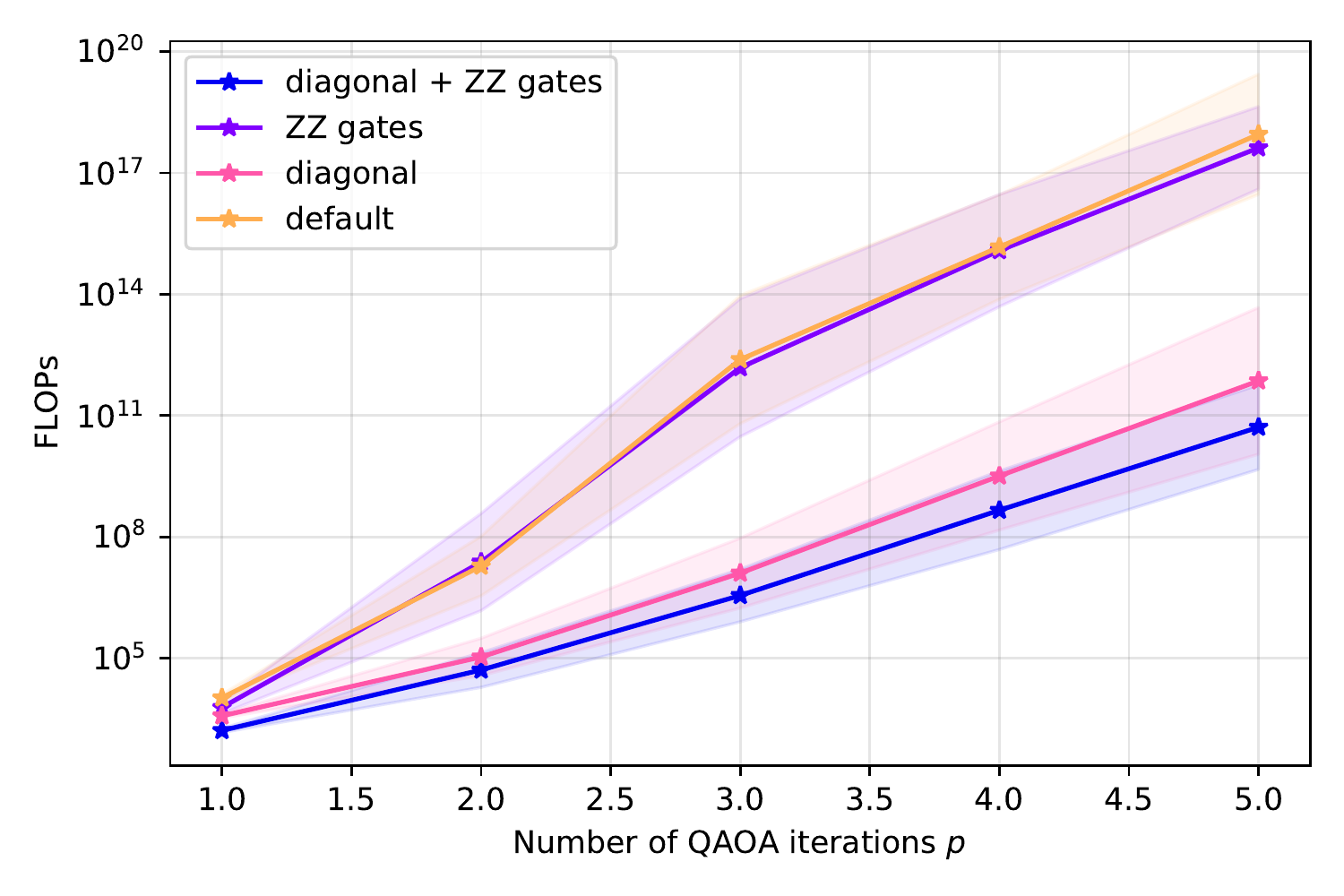}
\caption{
    Number of FLOPs to calculate a single amplitude
    of QAOA ansatz state for MaxCut using a different 
    number of QAOA iterations.
    Each line shows a combination of 
    optimization techniques, 
    with ``diagonal + ZZ gates" being the most advanced one. \shadeddescr
}
\label{fig:diag_costs}
\end{figure}

\begin{figure}
\centering
\includegraphics[width=\linewidth]{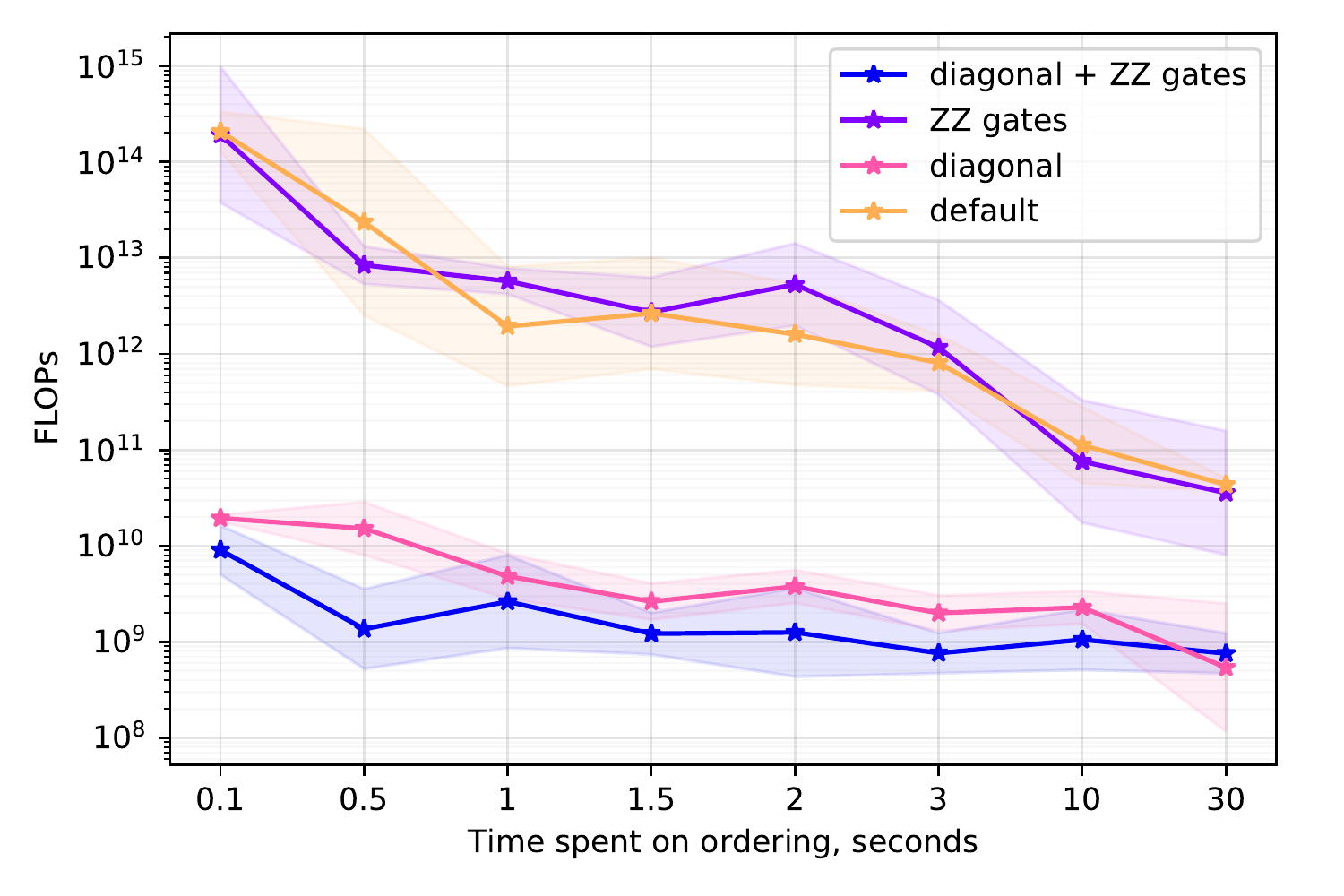}
\caption{
    Number of FLOPs to calculate a single amplitude
    of QAOA ansatz state with p=1 for MaxCut 
    on random 3-regular graphs with 160 nodes.
    \shadeddescr
}
\label{fig:diag_costs_opt_time}
\end{figure}

In this section we evaluate the simulation complexity of different
combinations of quantum circuits.
For this work we used the number of FLOPs required for contraction as the main metric of 
simulation complexity.
We used only the FLOPs metric because it is easy to use and both computational and memory requirements can be easily estimated from each other. The number of FLOPs can be estimated as $2^C$, where $C$ is a contraction width, or the number of dimensions of the largest 
intermediary tensor. Thus, one can safely assume that 
the maximum memory in bytes for the contraction will be $16*2^C$,
positing the size of a single complex number to be 16 bytes.

First, we selected \nseeds random 3-regular graphs,
for which we formulated the MaxCut problem by using QAOA.
Then, each graph was used to generate quantum circuits
that produce QAOA ansatz. There are two types of circuits:
one using ordinary decomposition of 2-qubit gates into three gates and
another using the $\hat{ZZ}$ gate simplification, as described in Section \ref{sec:zzgate}, where the complexity of the circuit is reduced.
For each quantum circuit, we constructed a tensor network using
two approaches: with diagonal simplification and without it, as 
described in  Section \ref{sec:diagonal_gates}. 
The tensor network was then sliced to produce 
the first amplitude of the ansatz state when all its indices are 
contracted.
We then used the \verb|rgreedy| algorithm from the QTensor package with 
\verb|n_repeats=10| and \verb|temp=0.02| to obtain a contraction sequence, which we used to
estimate the number of FLOPs required for the contraction.
Dependence of the contraction complexity from the contraction ordering time is shown in Figure \ref{fig:diag_costs_opt_time}.
The experiments in this paper aim to demonstrate the difference between
simulation of 
quantum circuits optimized using different techniques, rather than absolute values
of the simulation cost. Hence, we pick relatively modest ordering algorithm parameters that result in 1 second of ordering time.

Further analysis of contraction complexity is shown in Figure \ref{fig:diag_costs}. In this figure, the number of FLOPs
for simulating a single amplitude 
of QAOA ansatz circuit versus the number of QAOA iterations $p$ is shown. The data in the plot is evaluated for five random
3-regular graphs. From the figure one can see that diagonal gates and $\hat{ZZ}$ gate optimization techniques dramatically reduce the number of FLOPs compared with the original unoptimized tensor network graphs.

\begin{figure}
\centering
\includegraphics[width=\linewidth]{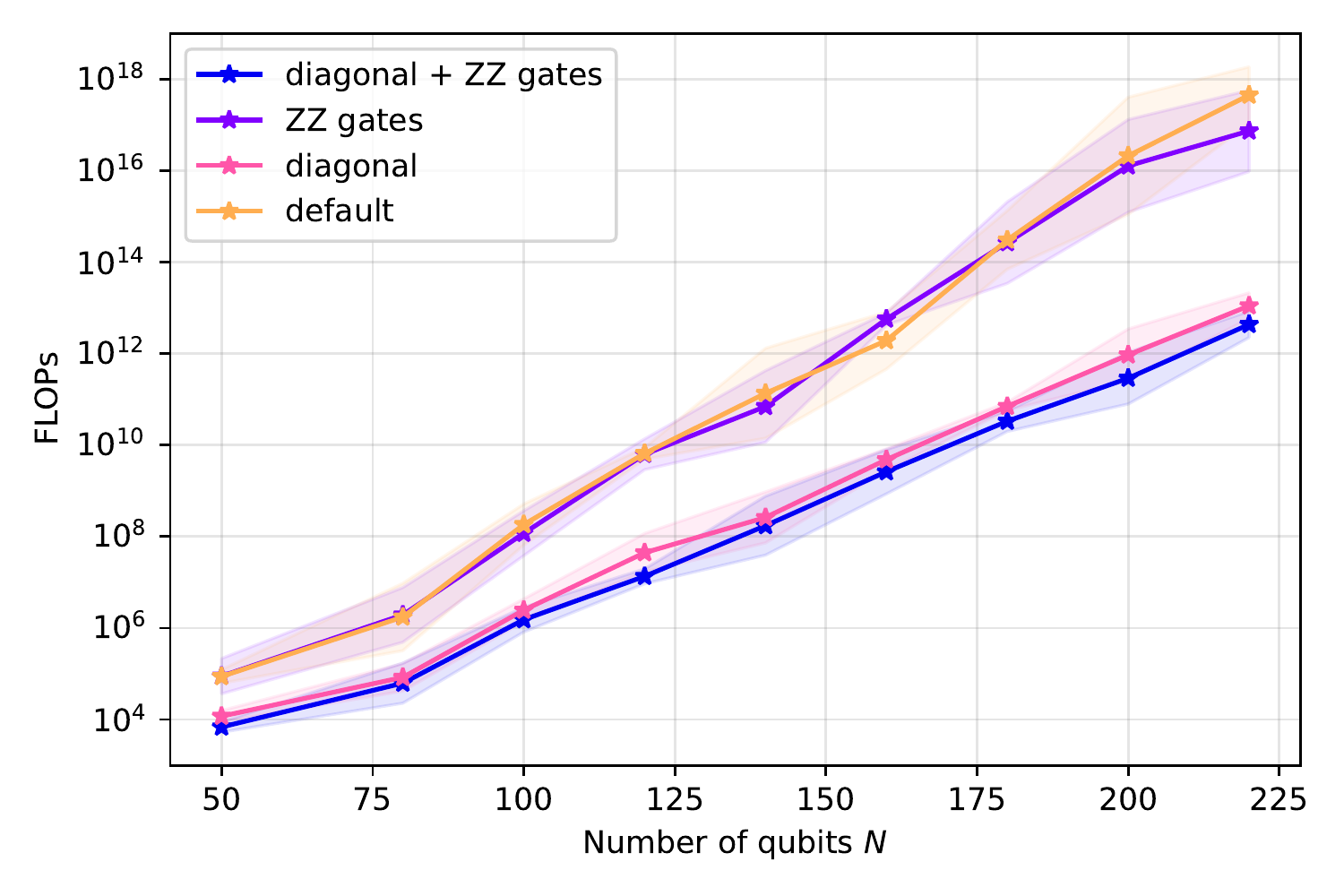}
\caption{
    The number of FLOPs to calculate a single amplitude
    of QAOA ansatz state for MaxCut on 
    random 3-regular graphs of different sizes.
    The number of nodes in the graph corresponds to 
    the number of qubits in the quantum circuit.
    Each line shows a combination of 
    optimization techniques, 
    with ``diagonal + ZZ gates" being the most advanced one. \shadeddescr
}
\label{fig:diag_costs_vs_n}
\end{figure}

We also analyzed how the number of FLOPs is correlated with the size of the circuit in terms of the number of qubits $N$. Figure \ref{fig:diag_costs_vs_n} shows the number of FLOPs
for ansatz simulation of QAOA MaxCut on 3-regular graphs of 
different sizes and fixed p=1.
We note that improvement in performance grows with the size of a 
circuit, showing that the diagonal gate simplification
proportionally reduces the contraction width $C$, 
while the $\hat{ZZ}$ gates simplification 
produces a similar amount of improvement for all sizes.

From the analysis of  Figures \ref{fig:diag_costs}, \ref{fig:diag_costs_opt_time}, and  \ref{fig:diag_costs_vs_n}, it is surprising to find out that using only $\hat{ZZ}$ gate optimization is not enough to get significant
computational and memory savings. We note that using the diagonal gates is what really provides dramatically better results, especially when  combined with gate optimization.

The simulation of quantum circuits is usually a memory-bound
task because of memory requirements to store an intermediate tensor. The optimization techniques described in this paper allow simplification of a tensor network structure with $\hat{ZZ}$ and diagonal gates, which in turn helped us find the optimal tensor contraction sequence.
As a result, we extended the depth $p$ and the number of qubits $N$ of
the largest circuits
that is feasible to simulate using a laptop or a supercomputer, as 
shown in Tables \ref{table} and  \ref{table_n}.
A laptop is assumed to have 4 GB of memory,
and a supercomputer is assumed to have 800 TB of aggregated memory.

Our optimization techniques also can be  applied to the simulation of the QAOA energy expectation value. Thus, it would allow us to find parameters for even larger circuits with higher depth.

\section{Conclusions}

\begin{table}[]
    {
        \begin{tabular}{|l|l|l|l|}
        \hline
        & Default  & With diagonal & With ZZ gates and diagonal \\
        \hline
        Laptop           & 2        & 3       & 3        \\
        Supercomputer    & 2        & 4       & 5           \\
        \hline
        \end{tabular}
    }
    \caption{
    \label{table}
    The maximum number of QAOA iterations $p$ for
    which one can simulate a single amplitude of
    ansatz state for MaxCut on a 40-node
    random regular graph.
    }
\end{table}

\begin{table}[]
    {
        \begin{tabular}{|l|l|l|l|}
        \hline
        & Default  & With diagonal & With ZZ gates and diagonal \\
        \hline
        Laptop           & 118        & 156        & 160         \\
        Supercomputer    & 180        & 240       & 244           \\
        \hline
        \end{tabular}
    }
    \caption{
    \label{table_n}
    The maximum number of nodes of a 3-regular graph
    for which one can simulate a single amplitude of
    the MaxCut QAOA ansatz state.
    }
\end{table}

In this work, we implemented and demonstrated that $\hat{ZZ}$ gate optimization and diagonal gate techniques can lead to dramatic savings in terms of memory and computation requirements. As a result, we were able to simulate much larger QAOA circuits both in size and in depth, as shown in Tables \ref{table} and \ref{table_n}.

We were able to increase $p$ from 2 to 5 and the size of a QAOA circuit $N$ from 180 to 244 qubits on a supercomputer. These numbers were estimated without actually running on a supercomputer, but the laptop results have been actually verified.

The tensor network contraction method for simulation of quantum circuits is a  powerful approach that has the potential to simulate very large quantum circuits.  At the same time, however, a lot of improvements, simplifications, and approximations  can be used to improve the simulations and dramatically decrease memory and computational requirements. For example, in this work we sped up quantum simulations by up to 10 million times, as can be seen in Figure \ref{fig:diag_costs} in the ``diagonal+ZZ gates" curve versus ``default" curve for $p=5$. Moreover, there is room to speed up simulations by an even larger factor.
This work underscores how one needs to be careful when comparing tensor network simulations against classical solvers and quantum hardware for demonstration of quantum supremacy and advantage. Our work provides important insights into how various optimization techniques can speed up tensor network simulations and what other techniques can be used to achieve this goal.

\section*{Acknowledgments}
Danylo Lykov and Yuri Alexeev are supported by the Defense Advanced Research Projects Agency (DARPA) grant. Yuri Alexeev is also supported in a part by the Exascale Computing Project (17-SC-20-SC), a joint project of the U.S. Department of Energy’s Office of Science and National Nuclear Security Administration, responsible for delivering a capable exascale ecosystem, including software, applications, and hardware technology, to support the nation’s exascale computing imperative.
This material is based upon work supported by the U.S. Department of Energy, Office of Science, under contract DE-AC02-06CH11357.
This work used the resources of the Argonne Leadership Computing Facility, which is DOE Office of Science User Facility supported under Contract DE-AC02-06CH11357. We thank Cameron Ibrahim and Alexey Galda for insightful discussions.

\bibliographystyle{IEEEtran}
\bibliography{main}

\end{document}

%% file: inputs/graph_figures.tex
\begin{figure*}
    \centering
    \input{inputs/examle_qaoa}
    \caption{Quantum circuit that generates QAOA ansatz state for MaxCut problem on a 4-node complete graph.
    This widely used decomposition of QAOA into common set of basis gates is not 
    optimal for the classical simulation of the output state.} 
    \label{fig:circ}
\end{figure*}

\begin{figure*}
    \centering
    \includegraphics[width=0.9\textwidth]{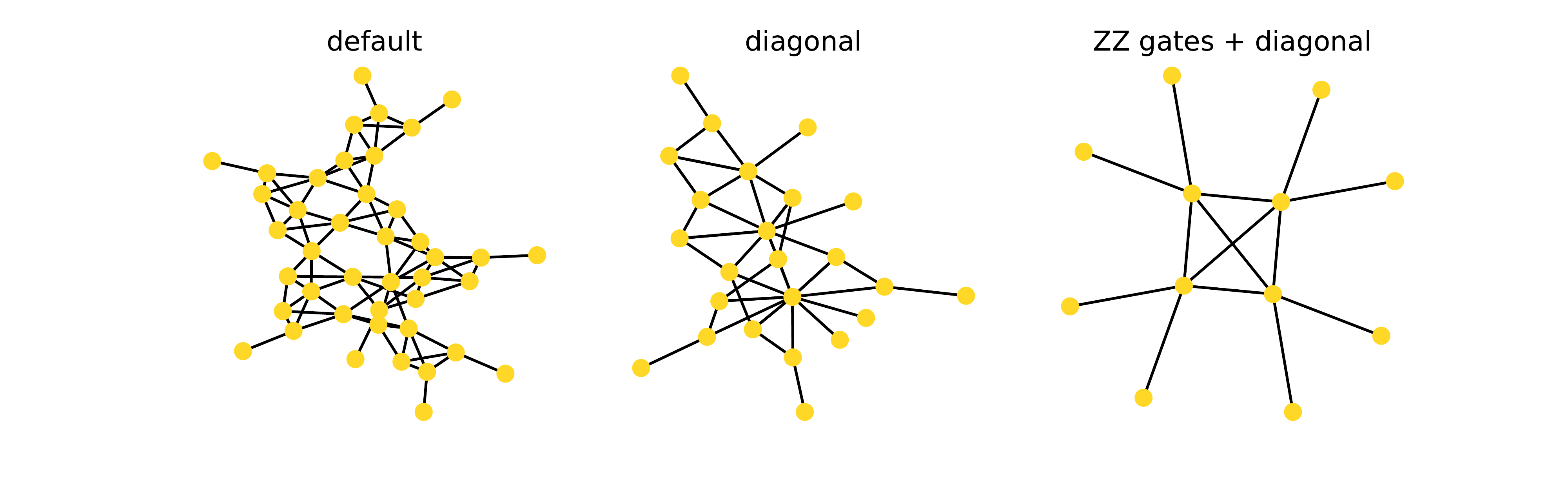}
    \caption{
    Line graphs of tensor networks for calculating QAOA ansatz state
    using different optimizations.
    ``Default" and ``diagonal" show line graphs of tensor network
    for the circuit shown in Figure \ref{fig:circ}, 
    using a full-matrix gates and diagonal gates approach, respectively. 
   ``ZZ gates + diagonal" is obtained by using
    the diagonal gates approach on a 
    simplified quantum circuit obtained by applying Equation \ref{eq:zz_gates_circ}.
    This figure demonstrates how improving the conversion
    of a quantum algorithm to a tensor network can
    reduce the complexity of the network, providing speedups
    for both finding contraction order and the contraction itself.
    }
    \label{fig:line_graphs}
\end{figure*}

%% file: inputs/examle_qaoa.tex
\Qcircuit @R=1em @C=0.75em {
 \\
 &\lstick{\text{0}}& \qw&\gate{\text{H}} \qw&\control \qw    &                                      \qw&\control \qw    &\control \qw    &                                      \qw&\control \qw    &\control \qw    &                                      \qw&\control \qw    &\gate{\text{H}} \qw    &\gate{\text{Z}^{2\beta}} \qw&\gate{\text{H}} \qw    &         \qw    &                                      \qw&         \qw    &                \qw    &                                      \qw&                \qw    &                \qw&                                      \qw&                \qw&\qw\\
 &\lstick{\text{1}}& \qw&\gate{\text{H}} \qw&\targ    \qw\qwx&\gate{\text{Z}^{2\gamma}} \qw&\targ    \qw\qwx&         \qw\qwx&                                      \qw&         \qw\qwx&         \qw\qwx&                                      \qw&         \qw\qwx&\control        \qw    &                                      \qw&\control        \qw    &\control \qw    &                                      \qw&\control \qw    &\gate{\text{H}} \qw    &\gate{\text{Z}^{2\beta}} \qw&\gate{\text{H}} \qw    &                \qw&                                      \qw&                \qw&\qw\\
 &\lstick{\text{2}}& \qw&\gate{\text{H}} \qw&         \qw    &                                      \qw&         \qw    &\targ    \qw\qwx&\gate{\text{Z}^{2\gamma}} \qw&\targ    \qw\qwx&         \qw\qwx&                                      \qw&         \qw\qwx&                \qw\qwx&                                      \qw&                \qw\qwx&\targ    \qw\qwx&\gate{\text{Z}^{2\gamma}} \qw&\targ    \qw\qwx&\control        \qw    &                                      \qw&\control        \qw    &\gate{\text{H}} \qw&\gate{\text{Z}^{2\beta}} \qw&\gate{\text{H}} \qw&\qw\\
 &\lstick{\text{3}}& \qw&\gate{\text{H}} \qw&         \qw    &                                      \qw&         \qw    &         \qw    &                                      \qw&         \qw    &\targ    \qw\qwx&\gate{\text{Z}^{2\gamma}} \qw&\targ    \qw\qwx&\targ           \qw\qwx&\gate{\text{Z}^{2\gamma}} \qw&\targ           \qw\qwx&         \qw    &                                      \qw&         \qw    &\targ           \qw\qwx&\gate{\text{Z}^{2\gamma}} \qw&\targ           \qw\qwx&\gate{\text{H}} \qw&\gate{\text{Z}^{2\beta}} 
 \qw&\gate{\text{H}} \qw&\qw\\
 \\
}

%% file: inputs/ZZ_replace.tex
\begin{equation}
\label{eq:zz_gates_circ}
\Qcircuit @C=.5em @R=.7em {
& \ctrl{1} & \qw & \ctrl{1} &\qw& \gate{H} & \gate{Z^{2\beta}} & \gate H \\
& \targ & \gate{Z^{2\gamma}} &\targ & \qw &
 \gate{H} & \gate{Z^{2\beta}} & \gate H 
}
\hspace{1em}
\raisebox{-1.2em}{=}
\hspace{1em}
\Qcircuit @C=.5em @R=.7em {
& \multigate{1}{ZZ} & \qw & \gate{X^{2\beta}} \\
& \ghost{ZZ} & \qw & \gate{X^{2\beta}}
}
\end{equation}

%% file: main.bbl
\begin{thebibliography}{10}
\providecommand{\url}[1]{#1}
\csname url@samestyle\endcsname
\providecommand{\newblock}{\relax}
\providecommand{\bibinfo}[2]{#2}
\providecommand{\BIBentrySTDinterwordspacing}{\spaceskip=0pt\relax}
\providecommand{\BIBentryALTinterwordstretchfactor}{4}
\providecommand{\BIBentryALTinterwordspacing}{\spaceskip=\fontdimen2\font plus
\BIBentryALTinterwordstretchfactor\fontdimen3\font minus
  \fontdimen4\font\relax}
\providecommand{\BIBforeignlanguage}[2]{{%
\expandafter\ifx\csname l@#1\endcsname\relax
\typeout{** WARNING: IEEEtran.bst: No hyphenation pattern has been}%
\typeout{** loaded for the language `#1'. Using the pattern for}%
\typeout{** the default language instead.}%
\else
\language=\csname l@#1\endcsname
\fi
#2}}
\providecommand{\BIBdecl}{\relax}
\BIBdecl

\bibitem{alexeev2019quantumworkshop}
Y.~Alexeev, D.~Bacon, K.~R. Brown, R.~Calderbank, L.~D. Carr, F.~T. Chong,
  B.~DeMarco, D.~Englund, E.~Farhi, B.~Fefferman, A.~Gorshkov, A.~Houck,
  J.~Kim, S.~Kimmel, M.~Lange, S.~Lloyd, M.~Lukin, D.~Maslov, P.~Maunz,
  C.~Monroe, J.~Preskill, M.~Roetteler, M.~Savage, and J.~Thompson, ``Quantum
  computer systems for scientific discovery,'' \emph{PRX Quantum}, vol.~2,
  no.~1, p. 017001, 2021.

\bibitem{de2007massively}
K.~De~Raedt, K.~Michielsen, H.~De~Raedt, B.~Trieu, G.~Arnold, M.~Richter,
  T.~Lippert, H.~Watanabe, and N.~Ito, ``Massively parallel quantum computer
  simulator,'' \emph{Computer Physics Communications}, vol. 176, no.~2, pp.
  121--136, 2007.

\bibitem{smelyanskiy2016qhipster}
M.~Smelyanskiy, N.~P. Sawaya, and A.~Aspuru-Guzik, ``{qHiPSTER}: the quantum
  high performance software testing environment,'' \emph{arXiv preprint
  arXiv:1601.07195}, 2016.

\bibitem{haner20170}
T.~H{\"a}ner and D.~S. Steiger, ``0.5 petabyte simulation of a 45-qubit quantum
  circuit,'' in \emph{Proceedings of the International Conference for High
  Performance Computing, Networking, Storage and Analysis}.\hskip 1em plus
  0.5em minus 0.4em\relax ACM, 2017, p.~33.

\bibitem{wu2019full}
X.-C. Wu, S.~Di, E.~M. Dasgupta, F.~Cappello, H.~Finkel, Y.~Alexeev, and F.~T.
  Chong, ``Full-state quantum circuit simulation by using data compression,''
  in \emph{Proceedings of the High Performance Computing,Networking, Storage
  and Analysis International Conference (SC19)}.\hskip 1em plus 0.5em minus
  0.4em\relax Denver, CO, USA: IEEE Computer Society, 2019.

\bibitem{wu2018amplitude}
X.-C. Wu, S.~Di, F.~Cappello, H.~Finkel, Y.~Alexeev, and F.~T. Chong,
  ``Amplitude-aware lossy compression for quantum circuit simulation,'' in
  \emph{Proceedings of 4th International Workshop on Data Reduction for Big
  Scientific Data (DRBSD-4) at SC18}, 2018.

\bibitem{wu2018memory}
X.-C. Wu, S.~Di, F.~Cappello, H.~Finkel, Y.~Alexeev, and F.~Chong,
  ``Memory-efficient quantum circuit simulation by using lossy data
  compression,'' in \emph{Proceedings of the 3rd International Workshop on
  Post-Moore Era Supercomputing (PMES) at SC18}, Denver, CO, USA, 2018.

\bibitem{quac}
\BIBentryALTinterwordspacing
(2020) {QuaC} (quantum in c) is a parallel time dependent open quantum systems
  solver. [Online]. Available: \url{https://github.com/0tt3r/QuaC}
\BIBentrySTDinterwordspacing

\bibitem{markov2008simulating}
I.~L. Markov and Y.~Shi, ``Simulating quantum computation by contracting tensor
  networks,'' \emph{SIAM Journal on Computing}, vol.~38, no.~3, pp. 963--981,
  2008.

\bibitem{pednault2017breaking}
E.~Pednault, J.~A. Gunnels, G.~Nannicini, L.~Horesh, T.~Magerlein,
  E.~Solomonik, and R.~Wisnieff, ``Breaking the 49-qubit barrier in the
  simulation of quantum circuits,'' \emph{arXiv preprint arXiv:1710.05867},
  2017.

\bibitem{boixo2017simulation}
S.~Boixo, S.~V. Isakov, V.~N. Smelyanskiy, and H.~Neven, ``Simulation of
  low-depth quantum circuits as complex undirected graphical models,''
  \emph{arXiv preprint arXiv:1712.05384}, 2017.

\bibitem{lykov2021large}
D.~Lykov, R.~Schutski, A.~Galda, V.~Vinokur, and Y.~Alexeev, ``Tensor network
  quantum simulator with step-dependent parallelization,'' \emph{arXiv preprint
  arXiv:2012.02430}, 2020.

\bibitem{qtensor}
D.~Lykov, ``{QTensor},'' \url{https://github.com/danlkv/qtensor}, 2021.

\bibitem{farhi2016quantum}
E.~Farhi and A.~W. Harrow, ``Quantum supremacy through the quantum approximate
  optimization algorithm,'' \emph{arXiv preprint arXiv:1602.07674}, 2016.

\bibitem{cichocki2016tensor}
A.~Cichocki, N.~Lee, I.~Oseledets, A.-H. Phan, Q.~Zhao, D.~P. Mandic
  \emph{et~al.}, ``Tensor networks for dimensionality reduction and large-scale
  optimization: {Part 1} -- low-rank tensor decompositions,'' \emph{Foundations
  and Trends{\textregistered} in Machine Learning}, vol.~9, no. 4-5, pp.
  249--429, 2016.

\bibitem{schutski2019adaptive}
R.~Schutski, D.~Lykov, and I.~Oseledets, ``An adaptive algorithm for quantum
  circuit simulation,'' \emph{arXiv preprint arXiv:1911.12242}, 2019.

\bibitem{detcher2013bucket}
\BIBentryALTinterwordspacing
R.~Dechter, ``Bucket elimination: {A} unifying framework for several
  probabilistic inference,'' \emph{CoRR}, vol. abs/1302.3572, 2013. [Online].
  Available: \url{http://arxiv.org/abs/1302.3572}
\BIBentrySTDinterwordspacing

\end{thebibliography}
